\documentstyle{mn}
\include{epsf}

\begin{document}

\def\spose#1{\hbox to 0pt{#1\hss}}
\def\lta{\mathrel{\spose{\lower 3pt\hbox{$\mathchar"218$}}
        \raise 2.0pt\hbox{$\mathchar"13C$}}}
\def\gta{\mathrel{\spose{\lower 3pt\hbox{$\mathchar"218$}}
        \raise 2.0pt\hbox{$\mathchar"13E$}}}

\title{Disc instability models for X-ray transients: evidence for evaporation
	and low $\alpha-$viscosity ?}

\author[K. Menou, J.-M. Hameury, J.-P. Lasota and R. Narayan]
{Kristen Menou$^{1,2,5}$\thanks{Present address: Princeton University, Department of Astrophysical Sciences, Princeton NJ 08544, USA, kristen@astro.princeton.edu}, Jean-Marie Hameury$^{3,5}$, Jean-Pierre
Lasota$^{4,5,2}$ \cr
and Ramesh Narayan$^{1,5}$
\\$^1$ Harvard-Smithsonian Center for Astrophysics, 60 Garden Street,
    Cambridge, MA 02138, USA,
\\ \ \ \  rnarayan@cfa.harvard.edu
\\$^2$ UMR 8629 du CNRS, D\'epartement d'Astrophysique Relativiste et de
        Cosmologie, Observatoire de Paris, Section de Meudon,
\\      F-92195 Meudon C\'edex, France
\\$^3$ UMR 7550 du CNRS, Observatoire de Strasbourg, 11 rue de l'Universit\'e,
        F-67000 Strasbourg, France, hameury@astro.u-strasbg.fr
\\$^4$ Institut d'Astrophysique de Paris, 98bis Boulevard Arago, F-75014 Paris,
        France, lasota@iap.fr
\\$^5$ Institute for Theoretical Physics, University of California, Santa
Barbara, CA 93106-4030, USA}

\maketitle

\begin{abstract}
We construct time-dependent models of accretion discs around black holes and
neutron stars. We investigate the effect of evaporating the disc inner regions
during quiescence on the predictions of the Disc Instability Model (DIM) for
these systems. We do not include irradiation of the disc in the models.

Removing the inner, most unstable parts of the accretion disc increases the
predicted recurrence times. However, DIMs with values of the viscosity
parameter $\alpha_{\rm hot} \sim 0.1$ and $\alpha_{\rm cold} \sim 0.02$
(typical of applications of the DIM to standard dwarf nova outbursts) fail to
reproduce the long recurrence times of SXTs (unless we resort to fine-tuning
of the parameters) independent of the evaporation strength. We show that
models with evaporation and a smaller value of $\alpha_{\rm cold}$ ($\sim
0.005$) do reproduce the long recurrence times and the accretion rates at the
level of the Eddington rate observed in outburst. The large difference between
the values of $\alpha_{\rm hot}$ and $\alpha_{\rm cold}$, if confirmed when
disc irradiation is included, suggests that several viscosity mechanisms
operate in these accretion discs.

For some choices of parameters our models predict reflares during the decline
from outburst. They are a physical property of the model and result from a
heating front forming in the wake of an initial cooling front and subsequent,
multiple front reflections. The reflares disappear in low--$\alpha$ models
where front reflection can not occur.
\end{abstract}


\section{Introduction}

Low Mass Black Hole Binaries (LMBHBs) and Low Mass Neutron Stars Binaries
(LMNSBs) are close binary systems in which a Roche-lobe filling, low mass
(main sequence or sub-giant) secondary star transfers mass to a compact
primary. All known LMBHBs and several LMNSBs are transients (see e.g. Tanaka
\& Lewin 1995, van Paradijs \& McClintock 1995, White, Nagase \& Parmar 1995);
these transient systems are known as soft X-ray transients (SXTs), or X-ray
novae. Their outbursts last from several weeks to several months during which
X--ray luminosities can approach the Eddington limit. The outbursts are
separated by long periods of quiescence lasting from one to tens of years (see
e.g. Chen, Shrader \& Livio 1997; hereafter CSL).

SXTs are in some respects similar to dwarf novae (DN), which exhibit outbursts
of 4 -- 6 magnitudes in optical. These outbursts typically last days and are
separated by periods of quiescence of a few weeks. DN are a subclass of
Cataclysmic Variables (CVs; see Warner 1995 for a review), close binary
systems similar to LMBHBs and LMNSBs, except for the compact primary which is
a white dwarf.

Is is widely believed that dwarf nova outbursts are due to a thermal-viscous
instability in the accretion disc of these systems. For effective temperatures
between about 5000~K and 8000~K, accretion discs in which viscosity is
described by the ``$\alpha-$prescription'' (Shakura \& Sunyaev 1973) are
thermally and viscously unstable because of large changes in the opacity when
hydrogen recombines. The disc instability model (DIM; see Cannizzo 1993b for a
review; Hameury et al. 1998 and Menou, Hameury \& Stehle 1999a for the most
recent version of the model) accounts for a number of general properties of
dwarf novae. Similarities between the properties of DN and SXTs led to the
suggestion that their outbursts have the same origin (van Paradijs \& Verbunt
1984; Cannizzo, Ghosh \& Wheeler 1982; Huang \& Wheeler 1989; Mineshige \&
Wheeler 1989).

Several important features of DN remain unexplained by the standard version of
the DIM, which assumes that the disc extends down to the white dwarf's
surface, that the mass transfer rate from the secondary star is constant, and
that effects of disc and secondary irradiation can be neglected (see e.g.
Lasota \& Hameury 1998; Lasota 1999; Warner 1998; Smak 1999, Hameury. Lasota
\& Warner 1999b for a discussion of these effects). This version cannot account
for the quiescent accretion rates onto the white dwarf (they exceed the model
predictions by more than two orders of magnitude), and for the very long
recurrence times of WZ~Sge--type systems (Lasota 1996a,b). The observed delay
between the rise in optical and in EUV (e.g. in SS Cyg; Mauche 1996) was also
long considered as a problem, but Smak (1998) recently argued that it can be
naturally reproduced by the DIM if appropriate outer boundary conditions for
the disc are used.

Similar difficulties, but more pronounced, appear for SXTs, since the standard
DIM fails to reproduce a number of essential properties of these systems. In
particular, an early attempt to explain the properties of BH SXT outbursts
with the DIM (Mineshige \& Wheeler 1989) was unable to reproduce the observed
long recurrence times (comparable to those of WZ~Sge systems) and the high
accretion rates required in outburst. In addition, the weak X-ray flux
detected from quiescent BH SXTs, which was mistakenly thought by some to be an
argument in favour of the DIM versus the competing mass transfer instability
model, is many orders of magnitude larger than any flux predicted by the DIM
\cite{l96a}.

In the DN case, the problems can been solved, at least in part, if a ``hole''
is present in the inner regions of the disc. The ``hole'' can either be caused
by evaporation of the disc \cite{mm94}, or due to the extended magnetospheric
cavity of the WD \cite{lp92,lkc99}. A model in which a ``hole'' is included
can account for both the EUV delay and the quiescent X-ray luminosity (if the
hole is filled with a tenuous, X--ray emitting gas; Meyer \& Meyer-Hofmeister
1994; Hameury, Lasota \& Dubus 1999a).

The case of WZ Sge requires additional modifications of the standard DIM. The
very long recurrence time of this system, and the large amount of mass
accreted during the outburst, can be reproduced either by using a value of the
viscosity parameter $\alpha_{\rm cold}$ several orders of magnitude lower than
in other DN (Smak 1993; Osaki 1995), or by making the ``hole'' sufficiently
large for the remaining disc to be globally stable. In the latter case, the
outbursts would have to be triggered by enhanced mass transfer (Lasota,
Hameury \& Hur\'e 1995; Hameury, Lasota \& Hur\'e 1997b). A marginally stable,
truncated disc, as proposed by Warner, Livio \& Tout \shortcite{wlt96}, is
equivalent to a globally stable disc from this point of view \cite{hlh97}. If
the truncated disc is globally stable, and $\alpha_{\rm cold}\sim 0.02$,
however, there is not enough mass in the disc to account for the outburst's
total energy, so that enhanced mass transfer is also required during the
outburst. Hameury et al. \shortcite{hlh97} find that if this enhanced mass
transfer is due to irradiation of the secondary, the model reproduces the
shape of the observed lightcurve.

The presence of ``holes'' in the accretion discs of quiescent SXTs seems
unavoidable if their outbursts are due to the thermal-viscous instability
(Lasota 1996a,b). In the BH case, such holes can only result from the
evaporation of the accretion disc since black holes are not magnetized.
Narayan, McClintock \& Yi (1996) proposed a model in which, as a result of
evaporation, the inner parts of the accretion flow in quiescent BH SXTs form
an advection-dominated accretion flow (ADAF; see also Lasota, Narayan \& Yi
1996; Narayan, Barret \& McClintock 1997). The model explains the observed
spectra and solves the problem of incompatibility between the accretion rates
predicted by the DIM and the observed flux.

Esin, McClintock \& Narayan (1997; see also Narayan 1996; Esin et al. 1998)
extended the idea of two-component accretion flows a step further. They showed
that the various spectral states of black hole X-ray binaries can be modeled
as a sequence of physical states with varying accretion rates and sizes of the
two components of the accretion flow. According to this model, in outburst,
the disc extends down to the black hole, while the inner disc is gradually
replaced by an ADAF as a system goes to lower luminosity levels. A gradual
evaporation of the inner disc during the decline is also qualitatively
consistent with the accretion geometry inferred by \.Zycki, Done \& Smith
(1999) from the X-ray reprocessing properties of several BH X-ray transients.

Hameury et al. (1997a) used the two component (inner ADAF + outer disc) model
to describe the quiescent state and the rise to outburst of the BH SXT GRO
J1655-40. The model predictions agree with existing constraints on the
spectrum of the system in quiescence, the observed 6 day delay between the
rise to outburst of the optical and X-ray fluxes, and the rise-times of these
two fluxes. Recently, Esin, Lasota \& Hynes (1999) showed that the whole,
unusual outburst can be described in the framework of the DIM thus validating
Hameury et al. (1997a) results.

In this paper, we consider the role of disc evaporation on the predictions of
the DIM for SXT outbursts. We shall refer to a DIM that includes evaporation
as a Truncated--Disc Instability Model (hereafter TDIM). Cannizzo (1998)
already constructed TDIMs of the BH SXT A0620-00 with a weak evaporation and
the prescription $\alpha=50(H/R)^{3/2}$ for the viscosity parameter.
Cannizzo's simulations show that models without evaporation lead to
unreasonably short recurrence times, and that evaporation acts to increase the
predicted recurrence times to values of several tens of years, as required.
This is expected, since evaporation removes the inner, most unstable parts of
the accretion disc. However, the outburst rise time predicted by Cannizzo's
model is an order of magnitude longer than the observed one. This is due to
the use of the prescription $\alpha=50(H/R)^{3/2}$ according to Lasota \&
Hameury (1998).

There are two additional reasons to go beyond Cannizzo's calculations. First,
his calculations do not allow the disc outer radius to vary with time. Smak
(1984, 1998; see also Hameury et al. 1998) showed that including this effect
is crucial for reliable DIM predictions of DN outbursts, and it is likely to
be important for SXT models as well. Second, in the models of Cannizzo
(1998), the effect of including evaporation and using a very small value of
$\alpha$ in the quiescent disc are not clearly separated. In particular, in
his work, it is unclear if evaporation alone can be held responsible for the
long recurrence times, if values of $\alpha$ similar to those inferred from
applications of the DIM to DN (typically $\alpha_{\rm hot} \sim 10^{-1}$ and
$\alpha_{\rm cold} \sim 0.02$) are used for the discs of SXTs. In our view,
this is an important question because the evidence for disc evaporation, even
indirect, is stronger than for any specific $\alpha-$prescription.

In \S~2, we recall some important observational properties of BH and NS SXTs,
later used to constrain our models. In \S~3, we first describe the numerical
truncated disc instability model used. We then show that, in the limit where
illumination can be neglected, TDIMs with values of $\alpha$ considered as
standard for DN discs are unable to reproduce the long recurrence times of
SXTs, no matter how strong the evaporation is. In \S~4, we present other
plausible models of SXTs. We find that a TDIM with a small value of
$\alpha_{\rm cold}$ ($\sim 5 \times 10^{-3}$) and strong evaporation does
reproduce the long recurrence times and the high maximum values of the
accretion rate in the discs of SXTs. We also present models in which the disc
of SXTs are globally stable and the outbursts are triggered by a slow
variation of the mass transfer rate of the secondary. In \S~5, we discuss
important consequences, limitations and possible extensions of this work. In
\S~6, we summarize our main results.

\section{Main observational characteristics of SXTs}

BH and NS X-ray Transients show a rather complex behaviour (see, e.g., Tanaka
\& Shibazaki 1996 and CSL for reviews). In this section, we identify several
important properties of these systems which may be considered as
characteristic of their class and therefore must be explained by a successful
model of SXT outbursts. We will also often compare SXTs with DN, since the
same underlying thermal-viscous instability is supposed to be responsible for
the outbursts in the two classes of systems.

As a guide, we use the observed properties of SS Cyg, Aql X-1 and A0620-00 as
indicative of the general properties of U Gem-type DN, NS SXTs and BH SXTs
respectively, keeping in mind that significant differences exist even between
two outbursts in a same system (e.g. at least four types of outbursts in SS
Cyg; Warner 1995).

\subsection{Timing properties}

The timescales of evolution of an unsteady disc are simple and robust
predictions of the DIM, which provide powerful tests of the model when
compared to the observations. This is because the time evolution of the disc
is due to the existence of a well defined limit cycle in the DIM, and because
the observed variability is `ready-to-interpret' (free of noise, systematic
errors, etc..).

Most of the properties of BH and NS SXTs mentioned in this section are taken
from the compilation of CSL. We identify several important timescales for
X-ray transients: the recurrence time $t_{\rm rec}$ between two successive
outbursts, the e-folding rise timescale $\tau_{\rm rise}$ and the total rise
time $t_{\rm rise}$ of an outburst (which can be wavelength-dependent), the
e-folding decay timescale $\tau_{\rm decay}$ (which is wavelength dependent),
and the total duration of the outburst $t_{\rm dur}$. In our view, a model
which fails to reproduce {\em all} these characteristic times is not fully
satisfactory. None of the previous models of SXTs passed this test (e.g.
Mineshige \& Wheeler 1989; Cannizzo 1998).

We focus our discussion on a sample of SXTs with clearly identified primaries
(see Garcia et al. 1998): the BH systems have firm dynamical lower limits on
the mass of the primary, and the NS SXTs showed type I X-ray bursts. We also
focus on systems which have exhibited standard Fast Rise Exponential Decay
(FRED) lightcurves. This excludes the long period systems GRO~ J1655-40 and
V~404~Cyg and reduces the sample to four BH SXTs (A0620-00, GS 2000+25, GRS
1124-683=Nova Mus 91 and GRO J0422+32 -- ``best examples of FREDs" according
to CSL) and two NS SXTs (Cen X-4 and Aql X-1; here we chose only their
outbursts of the FRED type : it is not clear how the standard DIM could
account for the other types of lightcurves observed in SXTs).

\subsubsection{Recurrence times}

Aql X-1 experiences an outburst every year approximately, while the recurrence
time of Cen X-4 is $\sim 10$ years. A0620-00 experienced two outbursts
separated by $58$ years (note that no X-ray data are available for the
presumed first outburst). Most BH SXTs did not experience a second outburst
since their discovery by X-ray satellites, so that there is good evidence for
the recurrence times of BH SXTs generally exceeding $20-30$ years. As
mentioned above, SXT recurrence times are similar to those of WZ Sge-type DN.
For comparison, the recurrence time of SS Cyg is $t_{\rm rec} \sim 50$ days
(e.g. Cannizzo \& Mattei 1992) and the recurrence time of other U Gem type DN
is generally of the order of one month (e.g. Szkody \& Mattei 1984).

\subsubsection{Rise times}

The characteristic X-ray rise timescales ($\tau_{\rm rise}$) of FRED-type SXTs
vary from 0.3 to 2 days (for the period of fastest flux increase -- CSL). The
total rise time $t_{\rm rise}$ from quiescence to outburst peak is difficult
to estimate because it is instrument dependent, and because the coverage of
the rising phase is, for obvious reasons, rather poor. For SXTs, $t_{\rm
rise}$ is $\sim$ 5 to 10 days in X--rays, or perhaps longer. For comparison,
the total optical rise time in SS Cyg is $t_{\rm rise} \sim$ 3 days (for a
`type B' outburst - see below). The optical rise time of several U Gem-type DN
in the sample of Szkody \& Mattei (1984) is $t_{\rm rise} \sim 2$ days.

In several DN, there is evidence for at least two types of outbursts. `Type A'
outbursts correspond to FREDs and are usually interpreted as `outside-in'
outbursts, i.e. outbursts triggered in the outer regions of the disc (Smak
1984). This interpretation seems, in a few cases, consistent with the observed
delay between the rise to outburst of optical and EUV (or UV) fluxes (Warner
1995). `Type B' outbursts have a more symmetric shape with a slower rise
phase, and are usually interpreted as `inside-out' outbursts (Smak 1984; see
also Menou et al. 1999a).

In the case of SXTs, there is no observational evidence for outside-in
outbursts. The delay between the rise of the optical and X-ray fluxes
(observed only during a non-FRED outburst of the BH SXT GRO J1655-40) {\em is
not} an indication of an outside-in outburst if the disc is truncated. Indeed,
Hameury et al. (1997a) obtain correct rise times and the X-ray/optical delay
for an inside-out outburst. In this case the `inside' corresponds to the inner
edge of the truncated disc, far from the central object.

\subsubsection{Decay times}

The X-ray decay timescales $\tau_{\rm dec}$ of SXTs with FRED light-curves are
typically 25 - 40 days (CSL). The optical decay timescales of BH SXTs are
$\sim$ 50 to 200 days, while the optical decay timescales of NS SXTs are
$\sim$ 10 to 30 days (but the available information is poor). We note that
long decay timescales are also observed in WZ Sge--type DN (Kuulkers, Howell
\& van Paradijs 1996; Kuulkers 1998).

\subsubsection{Duration times}

The reported duration times for the FRED--type outbursts of BH SXTs range from
170 to 260 days in X-rays (and even longer in optical), while $t_{\rm dur}
\sim 70$ days in X-rays for FRED--type outbursts of NS SXTs. The typical
duration time of U~Gem--type DN in the sample of Szkody \& Mattei (1984) is
$t_{\rm dur} \sim 10$ days (optical).

One has to remember, however, that the reported durations for SXTs depend on
the detection threshold of the instrument and that the coverage of some
outbursts is incomplete; sometimes the reported duration is only a lower limit
on the true duration. In some cases, the decay timescale $\tau_{\rm dec}$ may
be more appropriate than the duration time $t_{\rm dec}$ for comparison of the
model predictions with the observations.

\subsection{Other important observational constraints}

An estimate of the maximum luminosity reached in outburst by a SXT depends on
the distance to the system and, when compared to the Eddington luminosity, on
the mass of the primary. According to Garcia et al. (1998), the NS SXTs
considered here approximately reached the Eddington luminosity at outburst
peak, as did some of the BH SXTs. According to CSL, the maximum luminosities
reached are closer to 10\% of the Eddington luminosity, but their values could
be underestimated. For example, CSL obtain for Nova Muscae 1991 a peak
luminosity (0.4 -- 10 keV) of 0.08 in Eddington units, whereas according to
Esin et al. (1997), who used the entire available spectral information, the
peak luminosity in this system was almost exactly Eddington. In our view, this
means that models of SXTs have to be able to reach the Eddington accretion
rate. The Eddington accretion rate is defined here as \begin{equation} \dot
M_{\rm Edd}=1.39 \times 10^{18} \left( \frac{M_1}{M_{\odot}} \right)~{\rm
g~s^{-1}}, \end{equation} the rate at which an accretion flow around a central
object of mass $M_1$ reaches the Eddington luminosity for a standard $10\%$
radiative efficiency. The Eddington luminosity, however, seems to be the
maximum luminosity for SXTs. Combined with outburst durations this implies
that the mass accreted during outbursts by the central body is $\sim$ a few
$10^{24}$ g (roughly the same as accreted onto the white dwarf during WZ Sge
eruptions). This constraint must be satisfied by SXT models as well.

Observations of broadened $H_{\alpha}$ emission lines provide an upper limit
on the value of the disc inner radius $R_{\rm in}$ in quiescence for several
BH SXTs (e.g. Orosz et al. 1994; Narayan et al. 1996; Menou, Narayan \& Lasota  
1999b). In most systems, the constraints indicate a value of $R_{\rm in}$ less
than $10^4$ to a few $10^4 R_S$, where $R_s=2.95 \times 10^5 (M_1/M_\odot)$ cm
is the Schwarzschild radius of the black hole primary of mass $M_1$. The
limits provide a useful constraint on the strength of evaporation to be
included in the TDIMs of SXTs (see below).

For some systems, other constraints on $R_{\rm in}$ exist at various stages of
the outburst from the modeling of X-ray reflection properties (\.Zycki et al.
1998), as well as spectral fits which involve $R_{\rm in}$ as a
parameter (e.g. Sobczak et al. 1999). However, these two methods are model
dependent and somewhat controversial, because the relevant spectral features
can be modeled differently (M. Nowak, private communication). We choose to
ignore them here.

\section{``Standard'' truncated disc instability models of SXTs}

In this section, we investigate a ``standard'' version of the TDIM for SXTs.
As mentioned above, by ``standard", we mean a model with values of the
viscosity parameter $\alpha$ that are usually considered as standard for DN
discs: $\alpha_{\rm hot} \sim 0.1$ and $\alpha_{\rm cold} \sim 0.02$.
Non-standard models, in particular models using significantly smaller values
of $\alpha_{\rm cold}$, are considered in \S~4.

\subsection{Numerical specifications}

\subsubsection{General}

We use the numerical code described in Hameury et al. (1998) to simulate the
evolution of discs around NSs and BHs. The equations of conservation of mass,
angular momentum and energy for a thin Keplerian disc are solved on an
adaptive grid which resolves narrow structures in the disc (transition fronts
in particular). A fully implicit numerical scheme is used for time evolution,
to avoid any Courant-type limiting condition on the timestep of integration.

A grid of disc vertical structures, which give the local cooling rate of the
disc as a function of its surface density, central temperature and the
vertical gravity is precalculated before running the disc evolution. The
effect of changing the mass of the central object (from a NS SXT to a BH SXT)
enters the time-dependent calculations both through the radial conservation
equations solved, and through the grid of vertical structures with different
gravities.

The outer radius of the disc is allowed to vary with time under the action of
the torque applied by the secondary star on the outermost regions of the
disc. Including this proper boundary condition (as opposed to a rigid
boundary condition with a fixed value of the outer radius) has been found to
drastically affect the predictions of the DIM for DN discs (see Hameury et al.
1998 for details). The effect of including this proper boundary condition for
discs of SXTs is discussed in \S~3.2.1.

\subsubsection{Formation of a hole in the disc}

The formation of a hole in the discs of SXTs is probably determined by the
physics of the evaporation of gas from the disc, possibly into an ADAF. This
physics is poorly understood. Several mechanisms have been proposed in the
literature to explain the transition (see, e.g., Meyer \& Meyer-Hofmeister
1994; Liu, Meyer, Meyer-Hofmeister 1997; Narayan \& Yi 1995; Honma 1996;
Shaviv, Wickramasinghe \& Wehrse 1999), but none of these models has reached a
sufficient level of detail to allow robust quantitative predictions.

Disc evaporation has already been included in the calculations of Hameury et
al. (1997a) and Hameury et al. (1999a). In these studies, a `reasonable' formula
was used (close to the one proposed by Meyer \& Meyer-Hofmeister 1994) and
evaporation was included in the numerical models as a sink of matter in the
mass conservation equation (see also the models with evaporation of Cannizzo
1998).

Here, we chose a slightly different approach. We assume an {\sl ad hoc} law
for the rate of evaporation as a function of radius, as in Hameury et al.
(1997a; 1999a), but in addition we assume that evaporation acts only in a very
narrow region at the inner edge of the disc. This is a good approximation if
the rate of evaporation of the disc material is a steep function of radius.

To be more specific, the inner radius of the disc, $R_{\rm in}$, varies
according to:
\begin{equation}
\dot M_{\rm disc}(R_{\rm in})=\dot M_{\rm evap}(R_{\rm in}),
\label{eq:defrinevap}
\end{equation}
where $\dot M_{\rm disc}(R_{\rm in})$ is the time-dependent accretion rate at
the inner edge $R_{\rm in}$ of the disc, and $\dot M_{\rm evap}(R)$ is the
evaporation law with an arbitrary functional form. This evaporation law is
chosen so that, in outburst, $\dot M_{\rm evap}(R)$ is much less than the
accretion rate in the disc at any radius, and the disc is fully extended. The
radius of truncation of the disc in quiescence depends on the functional form
of $\dot M_{\rm evap}(R)$ and the profile of $\dot M_{\rm disc}$ in the disc.
We find that this numerical recipe is much more stable than that used in
Hameury et al. (1997a), while it gives very similar results for the evolution
of $R_{\rm in}$ and the outburst cycles in general.

Following Hameury et al. (1997a), we could have assumed that the disc is
evaporated into an ADAF which accretes everywhere at the maximum rate allowed,
$\dot M^{\rm ADAF}_{\rm crit}(R)$, and take for the evaporation law $\dot
M_{\rm evap}(R) = \dot M^{\rm ADAF}_{\rm crit}(R)$. Recently, however, the
estimates of $\dot M^{\rm ADAF}_{\rm crit}(R)$ have been revised to higher
values (Esin et al. 1997; Menou et al. 1999b). Using these recent estimates
for the evaporation law (i.e. a stronger evaporation than, e.g., in Hameury et
al. 1997a) would imply that truncated discs are globally stable (and
stationary) for all mass transfer rates below $\sim 10^{17}$ g s$^{-1}$ (this
is demonstrated by the Fig.~5 of Menou et al. 1999b), whereas typical SXT
mass--transfer rates are much lower: $\sim 10^{15} - 10^{16}$ g s$^{-1}$ (van
Paradijs 1996; CSL; Menou et al. 1999b). In addition, for mass transfer rates
$\le 10^{17}$ g s$^{-1}$, the stable discs would be truncated at radii $\ge 3
\times 10^4$ $R_s$, which is inconsistent with the upper limits derived from
$H_{\alpha}$ emission lines (\S~2.2).

To circumvent these difficulties, and ensure that discs in our models
experience outbursts even at lower mass transfer rates, $\dot M_{\rm evap}(R)$
must be less than $\dot M_{\rm crit}^{\rm ADAF}(R)$. We use the following
evaporation law:
\begin{equation}
\dot M_{\rm evap}(R)=\frac{0.08 \dot M_{\rm Edd}}{\left( \frac{R}{R_s}
\right)^{1/4} + {\cal E}\left( \frac{R}{800 R_s} \right)^2}.
\label{eq:defmdotevap}
\end{equation}
where ${\cal E}$ is a constant ensuring that a maximally truncated disc is
still unstable. The power laws and scalings in Eq.~(3) are such that the
profile of $\dot M_{\rm evap}(R)$ is initially slowly decreasing with radius,
and then decreases much more steeply beyond a radius $\sim 800 R_s \times
{\cal E}^{-1/2}$. In all the models with evaporation described below, we take
${\cal E}=20$ (except in section 4.1, in which ${\cal E}=30$). This corresponds
to a situation where the evaporation rate during quiescence is significantly
smaller than the mass transfer rate from the secondary, thus ensuring that the
disc is not marginally unstable. 

This expression for the evaporation law is {\sl ad hoc} but possesses several
advantages. First, it is a continuously decreasing function of radius, which
seems quite natural, and is appropriate for the numerical implementation of
evaporation described above. Second, the maximum value of $\dot M_{\rm
evap}(R)$, as defined in Eq.~(\ref{eq:defmdotevap}) when $R \sim R_S$, is
consistent with the high value required in the ADAF models of Esin et al.
(1997) to fit the high luminosity states of the BH SXT Nova Muscae 91. Third,
it leads to disc truncation radii in quiescence which are smaller than the
upper limits inferred from the broadening of $H_{\alpha}$ emission lines. The
resulting discs are unsteady and experience limit cycles for reasonable mass
transfer rates ($\sim 10^{15}$ g s$^{-1}$), as required for the DIM to be the
explanation of SXT outbursts.

Naturally, one could worry that the predictions of the time-dependent models
depend on the evaporation law used. This is true about the strength of
evaporation, but we find that our conclusions do not crucially depend on the
precise shape of the evaporation law. We discuss the effects of changing the
evaporation law on the model predictions in \S~5.

Note that in the models presented here, the presence of a tenuous region in
the inner accretion flow, presumably an ADAF, is only taken into account
through disc evaporation (i.e. as a sink of mass): we neglect any other
physical effect that this inner region of the accretion flow could have on the
disc evolution. Disc irradiation by the central X-ray source during high
luminosity states is also neglected, although it is a potentially important
effect (see \S~5 for a discussion). For simplicity, we use the same
evaporation law for BH and NS SXTs, except that $\dot M_{\rm Edd}$ and $R_S$
are scaled with the actual primary mass $M_1$.

\subsection{Results}

\begin{figure}
\epsfxsize=\columnwidth
\epsfbox{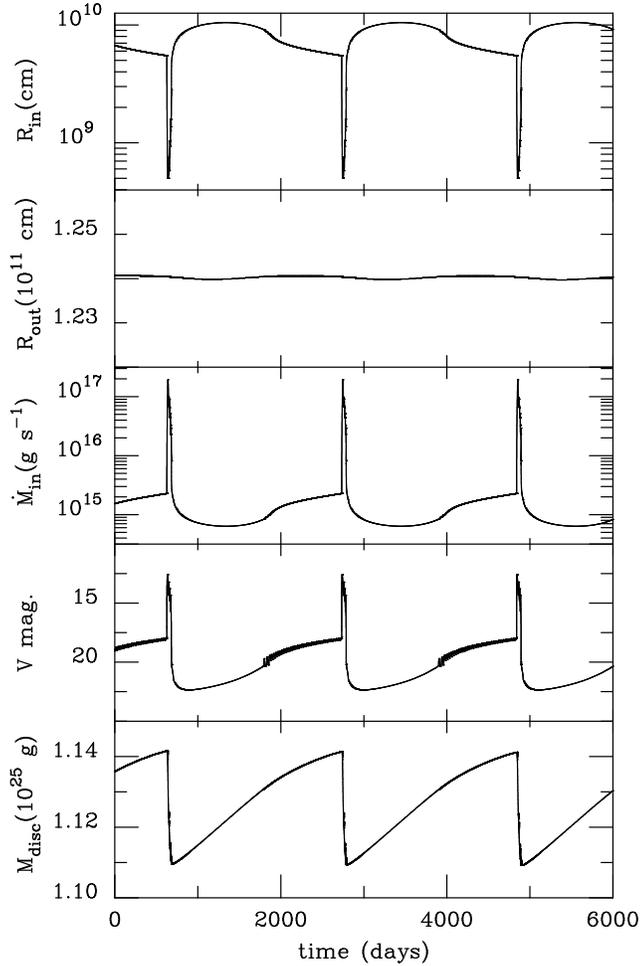}
\caption{Evolution of the disc mass $M_{\rm disc}$, the V band magnitude of
the disc, the accretion rate at the disc inner edge $\dot M_{\rm in}$, the
disc outer radius $R_{\rm out}$ and the disc inner radius $R_{\rm in}$ in a
model with parameters relevant for the BH SXT A0620-00 (see text for details).
The recurrence time is $t_{\rm rec} \simeq 5$ years in this model with
$\alpha_{\rm cold}=0.02$.}
\label{fig:a0620tot}
\end{figure}

\begin{figure}
\epsfxsize=\columnwidth
\epsfbox{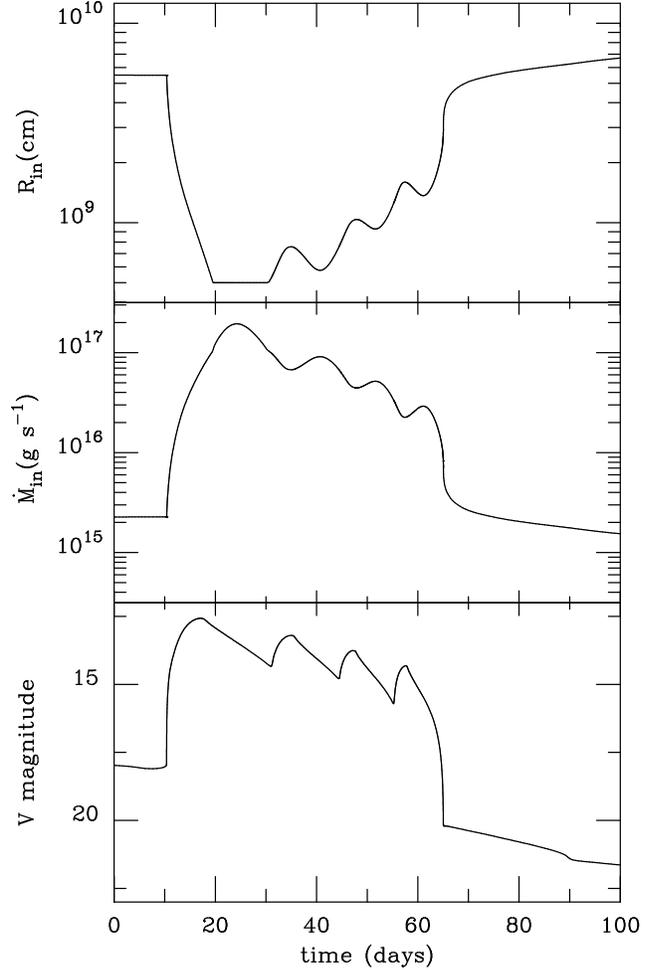}
\caption{Details on the evolution of the V band magnitude of the disc, the
accretion rate at the disc inner edge $\dot M_{\rm in}$ and the disc inner
radius $R_{\rm in}$ during an outburst of the model presented in Fig.~1. Note
the multiple reflares during the decline, as seen in the modulations of the V
band magnitude, $\dot M_{\rm in}$ and $R_{\rm in}$.}
\label{fig:a0620outburst}
\end{figure}

\begin{figure}
\epsfxsize=\columnwidth
\epsfbox{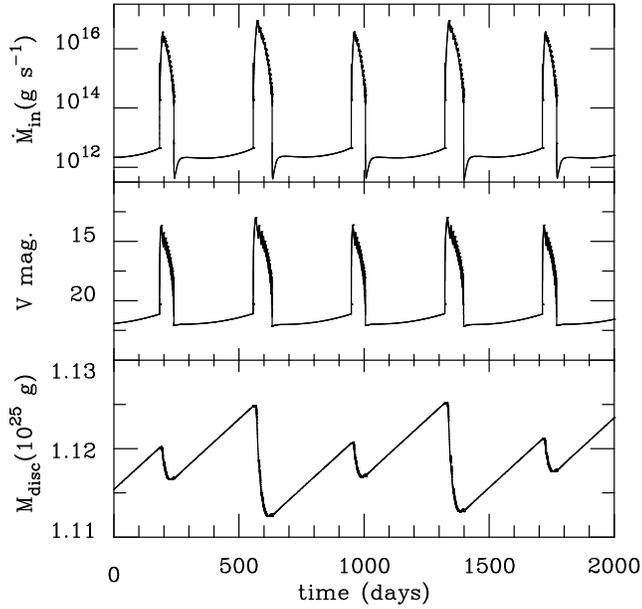}
\caption{Evolution of the total disc mass $M_{\rm disc}$, visual magnitude of
the disc and mass accretion rate onto the compact object $\dot M_{\rm in}$ in
a model with the same parameters as the one shown in Fig.~\ref{fig:a0620tot},
but without evaporation. The inner disc radius is fixed at $5 \times 10^8$ cm.
Here successive small and large outbursts are present.}
\label{fig:a0620cycle_noevap}
\end{figure}

\begin{figure}
\epsfxsize=\columnwidth
\epsfbox{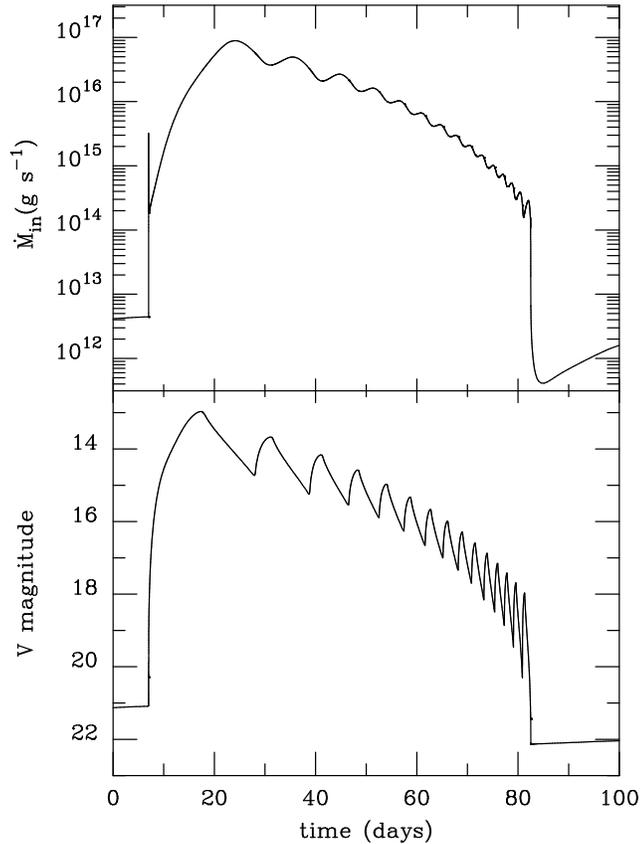}
\caption{Details of one of the outbursts of Fig.~\ref{fig:a0620cycle_noevap}}
\label{fig:a0620outburst_noevap}
\end{figure}

We present in this section two SXT models, one with parameters relevant for
the BH SXT A0620-00, and the other for the NS SXT Aql X-1. In both models,
$\alpha_{\rm hot}=0.1$, $\alpha_{\rm cold}=0.02$, the disc inner radius is
limited during the evolution to the minimum value $R_{\rm min}=5 \times 10^8$
cm (see \S~5 for a discussion of this numerical limitation) and
Eq.~(\ref{eq:defmdotevap}), scaled with $M_1$, is used for the evaporation law.

\begin{figure}
\epsfxsize=\columnwidth
\epsfbox{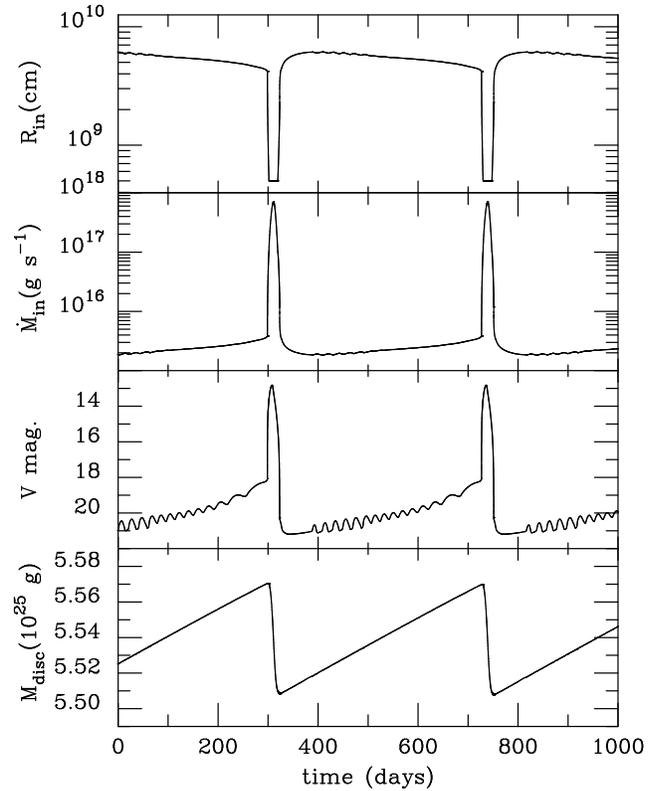}
\caption{Evolution of the total disc mass $M_{\rm disc}$, the V band magnitude
of the disc, the accretion rate at the disc inner edge $\dot M_{\rm in}$ and
the disc inner radius $R_{\rm in}$ for two outbursts in a model with
parameters relevant for the NS SXT Aql X-1. The recurrence time is $t_{\rm
rec} \simeq 1$ year in this model with $\alpha_{\rm cold}=0.02$. Here reflares
are absent.}
\label{fig:aqlx1outburst}
\end{figure}

The model of A0620-00 has the following parameters: primary mass $M_1=6
M_{\odot}$, time-averaged value of the disc outer radius $<R_{\rm out}>= 1.24
\times 10^{11}$ cm and mass transfer rate $\dot M_T=3 \times 10^{15}$ g
s$^{-1}$. The model of Aql X-1 has the following parameters: primary mass
$M_1=1.4M_{\odot}$, time-averaged value of the disc outer radius $<R_{\rm
out}>= 1.8 \times 10^{11}$ cm and mass transfer rate $\dot M_T=2 \times
10^{16}$ g s$^{-1}$. The main properties of these models are summarized in
Figures~\ref{fig:a0620tot} to~\ref{fig:aqlx1outburst}, and are discussed in
more detail below.

For comparison, we also show in Fig. \ref{fig:a0620cycle_noevap} and
\ref{fig:a0620outburst_noevap} results from a model with parameters relevant
for A0620-00, but without evaporation. Results are similar, except for a
significantly shorter recurrence time, a larger number of reflares, and the
existence of two type of outbursts. Successive small and large outbursts are
also predicted in models of dwarf novae (see e.g. Cannizzo 1993a; Menou et al.
1999a). During small outbursts, the outward propagating heating front does not
reach the outer disc radius. In the DIM, the inner regions of quiescent
accretion discs are more subject to instabilities than the outer ones, so that
the truncation of the inner disc has a stabilizing effect. This explains the
absence of `small' outbursts and the reduced number of reflares in models that
include evaporation (see below in \S~3.2.6).

The panel representing the total mass in Fig.~\ref{fig:a0620tot} shows that
the disc evolution is periodic and therefore relaxed (i.e. does not depend on
the arbitrary properties of the initial disc used at the beginning of the
simulation). This is true of every model discussed in this paper. Strictly
periodic outburst cycles are produced in SXT models, like for DN, while
observed cycles show, in general, only some regularity.

In the models presented here (as well as in all the other sections), the
outbursts are triggered in the inner regions of the truncated disc (inside-out
type). Such outbursts never start exactly at the disc inner edge, so that
there is always a heating front traveling some distance inwards. Its effect is
seen as a `spike' in the light-curve in Fig. 4, which lasts for less than one
minute and is therefore not observable. In a truncated disc, these short-lived
fronts are absent, because the innermost regions where they would propagate no
longer exist. Heating fronts only marginally reach the outer edge of the disc
in the model with parameters relevant for A 0620-00. In the model with
parameters relevant for Aql X-1, the disc is so extended that the heating
fronts never go beyond one third of the disc outer radius. This would of
course not be the case for shorter period systems such as GRO J0422+32.

In the following discussion of the rise times and decay times predicted by the
models, we will assume for simplicity that the X-ray emission from the system
is proportional to the accretion rate $\dot M_{\rm in}$ at the inner edge of
the disc. The real situation is likely to be more complicated, in particular
if some of the X-rays observed are emitted by an ADAF whose luminosity is not
simply proportional to $\dot M$ (in that case, the timescales could be even
shorter because the radiative efficiency increases with $\dot M$). Similarly,
the V magnitude shown are not supposed to directly represent the V magnitude
observable from a system but only the intrinsic optical luminosity variations
of the disc. In observed systems, the V band magnitude emission has also
contributions from the secondary, from reprocessing of X-ray photons by the
disc and possibly from an ADAF. We do not attempt to describe all these
processes here, and we concentrate on the basic properties of the models.

\subsubsection{Recurrence times}

As for DN discs, it is important to include a correct boundary condition at
the outer edge of the disc of SXTs. We find that the recurrence times
predicted by models in which $R_{\rm out}$ is allowed to vary with time are
typically a factor of a few (say 2-3) times shorter than those predicted if
$R_{out}$ is kept fixed. Indeed, when a heating front reaches $R_{\rm out}$,
mass is spread over a more extended region if $R_{\rm out}$ is allowed to
vary. Consequently the critical surface density $\Sigma_{\rm min}$ below which
a cooling front appears in the disc is reached by the disc earlier, less mass
is accreted during the outburst, and overall it takes less time for the disc
to reach the state of critical surface density $\Sigma_{\rm max}$ at which the
next outburst is triggered (Hameury et al. 1998). Therefore, all existing
models of BH SXTs with fixed values of $R_{\rm out}$ have systematically
overestimated the recurrence times $t_{\rm rec}$ by a factor of a few when the
heating front was able to reach the disc's outer edge. In the present paper,
we only consider models with varying $R_{\rm out}$.

The recurrence time in the model of A0620-00 is $t_{\rm rec} \sim 5$ years,
i.e. shorter than the known recurrence time by more than one order of
magnitude (\S~2.1.1). The recurrence time in the model of Aql X-1 is $t_{\rm
rec} \sim 1$ year, in very good agreement with the know recurrence time.

\subsubsection{quiescence}

During quiescence, all of the accreted mass is processed through the ADAF and
is therefore evaporated, so that the mass accretion rate shown in figs. 1 and 5
is the actual evaporation rate except when the disc radius reaches its minimum
value during outbursts. As can be seen, the evaporation rate remains
significantly smaller than the mass transfer rate from the secondary, varying
between 0.15 and 0.60 $\dot{M}_{\rm T}$ in the case of the A0620-00 model, and
between 0.10 and 0.15 $\dot{M}_{\rm T}$ in the Aql X-1 case. These are close to
the maximum evaporation rate for which the disc is unstable but they do not
imply fine tuning.

\subsubsection{Rise times}

In the model of A0620-00, once an outburst is triggered, it takes the disc
$\sim 9$ days to reach the compact object (the time required to go from $5
\times 10^8$ cm to $\sim 10^7$ cm, not included in our simulations, is small
because of the short timescales in the inner disc, so that we neglect it). It
also takes $\sim 14$ days for the accretion rate at the disc inner edge to
reach its maximum value. Overall, the total rise time in X-rays is therefore
$t_{\rm rise} \sim 5$ days, which is in reasonable agreement with the
available information on FRED-type light-curves. The total rise time in
optical is much shorter because the outbursts are triggered in a truncated
disc, i.e. effectively in the outer regions of the accretion flow. Our results
are similar to those of Hameury et al. (1997a) who addressed in more details
the issue of time delays between the rise in optical and X-ray.

The rise timescale for the period of fastest flux increase is $\tau_{\rm rise}
\simeq 4$ days in X-rays. This is somewhat longer than the observed rise
timescales of 0.3-2 days (\S~2.1.2). For completeness, we also give the rise
timescale in optical, $\tau_{\rm rise} \simeq 1$ hour, for which there is very
little available observational data (GRO J1655-40 being the only system in
which the rise in optical was observed; see Orosz et al. 1997). This timescale
is approximately the thermal time at the disc inner edge, since this is where
most of the optical light comes from at the beginning of an outburst. It is
not easily comparable to the observations, because of light dilution by both
the secondary and the ADAF.

In the model of Aql X-1 shown in Fig.~\ref{fig:aqlx1outburst}, it takes the
disc $\sim 3$ days to reach the compact object once an outburst is triggered.
The fact that this time is shorter than in the model of A0620-00 can be partly
attributed to a less efficient evaporation in this model (because $\dot M_{\rm
Edd}$ and $R_S$, in Eq.~[\ref{eq:defmdotevap}], scale with the mass $M_1$ of
the primary). It also takes $\sim 12$ days for the accretion rate at the disc
inner edge to reach its maximum value. Overall, the total rise time is
therefore $t_{\rm rise} \simeq 9$ days, which again is in reasonable agreement
with the available data on FRED-type light-curves.

The total rise time in optical is not so short: $t_{\rm rise} \simeq 9$ days.
This is because the disc is less strongly truncated than in the model of
A0620-00, which results in much more ``inside-out/symmetric'' outburst shapes
(e.g. Smak 1984). In addition, because the disc is large, the inner parts do
not contribute as much as in the case of A0620-00. The rise timescale for the
period of fastest flux increase is $\tau_{\rm rise} \simeq 1$ days in X-rays,
which is consistent with available observational data. It is $\tau_{\rm rise}
\simeq 1.7$ hours in optical.

\subsubsection{Decay and duration times}

In the model of Aql X-1, the total duration time of outbursts is $t_{\rm dur}
\simeq 25$ days. This is short compared to the duration $\sim 70$ days
reported in the literature for FRED-type outbursts. Again, we note that the
outbursts produced in this model are not really of the FRED-type, but have a
rather symmetric ``triangle'' shape.

The decay timescale early in the decline is $\tau_{\rm dec} \simeq
3.5$ days in X-rays and $\tau_{\rm dec} \simeq 4$ days in
optical. The timescales become smaller later on because the
light-curves steepen. These values are clearly shorter than the
observed values of 25-40 days in X-rays and 10-30 days in optical.

In the model of A0620-00, the total duration time of an outburst is $t_{\rm
dur} \simeq 55$ days. This is quite short compared to the typical duration
times for BH SXTs with FRED-type light-curves, $\sim170-260$ days in X-rays
and even more in optical. Note that $t_{\rm dur}$ would be shorter if reflares
did not occur (see below for a discussion of reflares).

Estimating the decay timescales is complicated by the presence of reflares in
this model. For completeness, we estimate two types of $\tau_{\rm dec}$. The
overall $\tau_{\rm dec}$ corresponds to the decay timescale for the global
light-curve, ignoring the ``fluctuations'' induced by the reflares. It is
$\tau_{\rm dec} \simeq 20$ days in X-rays and $\tau_{\rm dec} \simeq 22$ days
in optical. The actual decay timescales, in between two successive reflares,
are shorter: $\tau_{\rm dec} \simeq 9$ days in X-rays and $\tau_{\rm dec}
\simeq 11$ days in optical. We find it difficult to compare these timescales
with observed values, since observed light-curves do not resemble the
light-curves shown in Fig.~\ref{fig:a0620outburst} (see \S~5 for a discussion)

One should keep in mind that we did not take into account outer disc X--ray
irradiation which will increase significantly both the duration and the decay
timescale (Dubus 1999a,b).

\subsubsection{Outburst amplitude}

In the model of A0620-00, the maximum accretion rate reached at the inner edge
of the disc is substantially sub--Eddington ($\sim 0.015 \dot M_{\rm Edd}$;
Fig.~\ref{fig:a0620outburst}). This is far too small. In the model of Aql X-1,
on the other hand, the Eddington limit is reached without difficulty
(Fig.~\ref{fig:aqlx1outburst}).

The total amount of mass accreted during an outburst is also quite small in
the model of A0620-00. This is not surprising because the recurrence time is
more than 10 times shorter than observed, and not much mass is accumulated in
quiescence. On the other hand, the total accreted mass is comparable to what
is deduced from the observations in the case of Aql X-1. Again, this is not a
surprise since the recurrence time is about right.

\subsubsection{Reflares}

A striking feature of the models described here is the possible occurrence of
reflares during the decline of an outburst (Fig.~[\ref{fig:a0620outburst}]).
These reflares correspond to multiple reflections of cooling and heating
fronts during their propagation in the disc. For the specified set of
parameters, we find that they are present in models with a BH primary, but
absent when the primary is a NS. The number of reflares is reduced when the
disc is truncated because increasing the inner disc radius gradually removes
the regions of the disc where reflections occur.

Reflares are in fact a natural outcome of the DIM. They have been observed in
many simulations over the years, in particular in models of DN discs in which
a same value $\alpha_{\rm hot}=\alpha_{\rm cold}$ was used (e.g. Smak 1984).
In that case, a cooling front develops, propagates over some fraction of the
disc radial extent, is reflected into a heating front which goes back all the
way up to the disc outer edge. This is repeated indefinitely and leads to
small amplitude variations of the disc luminosity. This failure of the model
has been overcome by using a higher value of $\alpha$ for the hot ionized disc
than for the cold neutral disc, which results in large amplitude outbursts as
those observed.

The fundamental reason why a reflare occurs is because the surface density
just behind a cooling front reaches the critical value $\Sigma_{\rm max}$
sometime during its propagation. This happens typically when the front is
located at $R_{\rm front} \lta 10^{10}$~cm in our simulations. An outward
propagating heating front then appears which destroys the inner cooling front
as mass starts being accreted efficiently in the cooling region. Later on,
this heating front is reflected when the surface density just behind the front
reaches the critical value $\Sigma_{\rm min}$. A cooling front develops which
shuts off the outward transport of angular momentum at the origin of the
heating front propagation. A sequence of heating and cooling front reflections
corresponds to the multiple reflares observed in the simulations.

What was probably not previously realized is that the occurrence of reflares
depends not only on the values of $\alpha$ but also on the mass of the central
object, as shown by our simulations. We note that reflares were also found in
the simulations of the decline phase of BH SXT outbursts of Cannizzo, Chen \&
Livio (1995). Cannizzo et al. interpreted the reflares as numerical artifacts.
This is not the case in our models: reflares are a generic property of the DIM
and the TDIM for some range of masses and viscosity parameters.

One can rather easily see how the presence or absence of reflares depends on
the mass of the accreting object. As noticed by Menou et al. (1999a) the
surface density at the cooling front is
\begin{equation}
\Sigma\left(R_{\rm front}\right)= \Sigma_{\rm min}\left(R_{\rm front}\right),
\end{equation}
where $\Sigma_{\rm min}(R)$ is the `minimum' critical surface density in the
disc. Vishniac (1997) emphasized that the cold, outer regions of the disc
behind the cooling front are essentially frozen during the cooling front
propagation. Therefore, $\Sigma(R)$ behind the front is constant with time and
equal to $K(M_1, \alpha)\Sigma_{\rm min}$, where $K$ depends on the viscosity
parameter $\alpha$ and the mass of the accreting object $M_1$. According to
the disc models of Menou et al. (1999a), for a central mass $M_1=1.2$
M$_{\odot}$, $K\approx 4$, while for $M_1=7$ M$_{\odot}$, $K \approx 6-7$. The
ratio of the `maximum' to `minimum' surface densities is (Hameury et al. 1998)
\begin{equation}
{\Sigma_{\rm max} \left( \alpha_{\rm cold}=0.02\right) \over \Sigma_{\rm
min}\left(\alpha_{\rm hot}= 0.1 \right)} \approx 6.4,
\end{equation}
so that for $M_1=1.2$ M$_{\odot}$ the post-cooling-front surface density
$K\Sigma_{\rm min} < \Sigma_{\rm max}$, and no reflares are expected. For
$M_1=7$ M$_{\odot}$, the post-cooling-front surface density $K\Sigma_{\rm min}
\gta \Sigma_{\rm max}$ and it is not surprising that reflares are present.

To explain the dependence of $K$ on the mass of the accreting object, one can
use the cooling front model of Vishniac \& Wheeler (1996). Although some of
the predictions of this model were found to be somewhat inaccurate by Menou et
al. (1999a), we expect the scalings derived here to be valid. In this model
the front speed can be written as:
\begin{equation}
v_{\rm front} \propto \left({H \over R}\right)^{0.7}.
\end{equation}
From mass conservation the post-cooling front density can be written
as (see e.g. Vishniac 1997)
\begin{equation}
\Sigma_{\rm pf} \sim \Sigma_{\rm min} {v_r({\rm front}) \over v_{\rm front}},
\end{equation}
where $v_r({\rm front})$ is the speed of matter at the cooling front. Using
Vishniac \& Wheeler (1996) model, one obtains
\begin{equation}
{\Sigma_{\rm pf} \over \Sigma_{\rm min}} \propto M_1^{0.35}.
\end{equation}
This scaling, $K \propto M_1^{0.35}$, is in good agreement with the numerical
results of Menou et al. (1999a).

\subsubsection{Consequences}

Our models confirm the results of Cannizzo (1998) showing that a stronger
evaporation increases $t_{\rm rec}$ because it removes the inner, most
unstable parts of the disc. The recurrence times are also increased if the
mass transfer rate $\dot M_T$ in the model is reduced, since more time is then
needed to accumulate mass in the disc up to the critical surface density
$\Sigma_{\rm max}$ at which the next outburst is triggered. An exploration of
the parameter space of the models shows that it is not possible, however, to
reproduce recurrence times of ten to tens of years in TDIMs with `standard'
values of $\alpha$. This is because to obtain these long recurrence times, the
increase in the strength of evaporation or the reduction of $\dot M_T$
required lead to globally stable discs.

To demonstrate this, we computed a series of models of NS and BH SXTs with
fixed values of $R_{\rm in}$. The various values of $R_{\rm in}$ are
equivalent to various strengths of disc evaporation. For example, a model with
parameters relevant for A0620-00, with a fixed value of $R_{\rm in}=6 \times
10^{9}$ cm and $\dot M_T= 10^{16}$ g s$^{-1}$ has a recurrence time $t_{\rm
rec} \simeq 3$ years. This is slightly smaller than the recurrence time in the
model with evaporation shown in Fig.~\ref{fig:a0620tot}, in which $R_{\rm in}
\simeq 6 \times 10^{9}$~cm before the onset of an outburst, but $\dot{M}_T$ is
slightly smaller. This type of comparison shows that the main effect of
evaporation on the outburst cycles is to affect the radial extent of the disc
in quiescence, which in turn determines $t_{\rm rec}$.

\begin{figure}
\epsfysize=8.5cm
\epsfxsize=8.5cm
\begin{displaymath}
\epsfbox{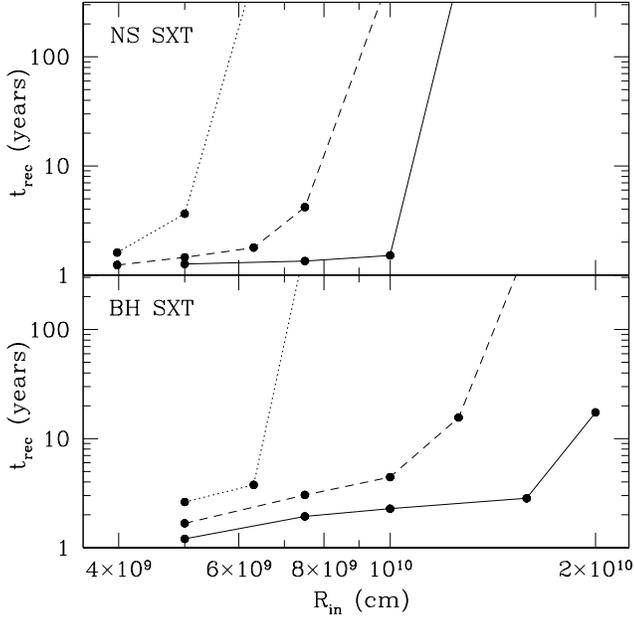}
\end{displaymath}
\caption{Recurrence times predicted by models of NS and BH SXTs (assuming
parameters relevant to Aql X-1 and A0620-00, respectively) for various values
of the disc fixed inner radius $R_{\rm in}$ ($\alpha_{\rm hot}=0.1$,
$\alpha_{\rm cold}=0.02$). The dotted, dashed and solid lines correspond to
mass transfer rates $\dot M_{T}=10^{16}, 3 \times 10^{16}$ and
$10^{17}$~g~s$^{-1}$, respectively. For both NS and BH SXTs, truncating more
the disc results in globally stable discs ($t_{\rm rec} \rightarrow \infty$)
instead of $t_{\rm rec} \geq 10-50$ years as expected. Models with arbitrary
long $t_{\rm rec}$ can be constructed by fine tuning a marginally stable disc
(e.g. the two upper points in the BH SXT panel), but these models do not have
any physical meaning (see text for details).}
\label{fig:trec}
\end{figure}

Figure~\ref{fig:trec} shows the recurrence times predicted for three mass
transfer rates ($\dot M = 10^{16}$, $3 \times 10^{16}$ and $10^{17}$ g
s$^{-1}$) and several values of the fixed inner radius $R_{\rm in}$. At a
given $\dot M_T$, the same trend is observed: the recurrence time initially
increases rather slowly with $R_{\rm in}$, but then the disc rapidly becomes
globally stable at some critical value of $R_{\rm in}$. For reasonable values
of $\dot M_T$, this critical value of $R_{\rm in}$ corresponds to recurrence
times well below ten to tens of years. It is possible to adjust the inner disc
radius so that the recurrence time will be arbitrary long (see e.g. the two
upper points in the lower panel of Fig.~\ref{fig:trec}) but the resulting disc
is marginally stable. Keeping it in this state for the entire quiescence
period requires a mass-transfer rate $\dot M_T$ strictly constant, which is
clearly unrealistic.

This result on short recurrence times is confirmed by an analytical estimate
of the recurrence times of SXTs.

An outburst starts when the surface density reaches the critical value
$\Sigma_{\rm max}$. In the case of outside-in (type A) outbursts, this happens
when enough mass has accumulated at the disc outer edge. If the accumulation
time at the outer edge is longer than the viscous time (as it is the case for
low accretion rates), matter transferred from the secondary diffuses inwards
and accumulates at shorter radii, giving rise to an inside-out (type B)
outburst. Since, for a given set of parameters, the recurrence time of a type
A outburst is shorter than the corresponding time for a type B outburst
($t_{\rm A} < t_{\rm B}$), we consider here only type B outbursts (in any case
the only type obtained in our models).

Following Smak (1993), the recurrence time of a type B outburst is
\begin{equation}
t_{\rm B} \approx \left( {\partial \ln \Sigma \over \partial t} \right)^{-1},
\end{equation}
which is the characteristic growth-time of a surface density contrast. From
the standard diffusion equation describing the evolution of $\Sigma$, one
obtains the relation between $t_{\rm B}$ and $\nu \Sigma$, where $\nu$ is the
kinematic viscosity coefficient. The relation between $\nu \Sigma$ and the
effective temperature $T_{\rm eff}$ is
\begin{equation}
\sigma_{\rm SB} T^4_{\rm eff}= \frac{9}{8} \frac{GM_1 \xi}{R^3} \nu \Sigma,
\end{equation}
where $\sigma_{\rm SB}$ is the Stefan-Boltzmann constant, $G$ is the
gravitational constant, $R$ is the distance from the central object of mass
$M_1$, and $\xi \approx 3 - 5$ is a constant factor that accounts for the
non-stationarity of the quiescent disc (Idan et al . 1999). Using the values
of $\Sigma_{\rm max}$ calculated by Hameury et al. (1998) and assuming a
constant effective temperature, which is a good approximation in quiescence
(Smak 1984; Hameury et al. 1998), one gets
\begin{eqnarray}
t_{\rm B} \approx 3 \left(\frac{\xi}{3}\right)\left(\frac{M_1}{M_{\odot}}
\right)^{0.62} \left( \frac{R}{10^{10}~{\rm cm}}\right)^{0.14}
\times \nonumber \\
\left( \frac{\alpha_{\rm cold}}{0.02} \right)^{-0.83}
\left(\frac{T_{\rm eff}}{3000\ {\rm K}}\right)^{-4}
\ & {\rm yr}.
\label{tb}
\end{eqnarray}
This characteristic time $t_{\rm B}$ is expected to be an upper limit to the
recurrence time (e.g. $T_{\rm eff}$ may be larger than 3000 K). Again, this
rough estimate is much less than tens of years. It also suggests that one way
to obtain long enough recurrence times is to reduce the value of $\alpha_{\rm
cold}$. Interestingly, we expect the maximum accretion rate reached by the
disc during outburst to increase as well if $\alpha_{\rm cold}$ is reduced.
This is because more mass will be present in the disc at the time an outburst
is triggered ($\Sigma_{\rm max} \propto \alpha_{\rm cold}^{-0.83}$; e.g.
Hameury et al. 1998). Models of SXTs with smaller values of $\alpha_{\rm
cold}$ are considered in the following section.

\section{``Non-standard'' truncated disc models of SXTs}

In the following, we present two non-standard truncated disc models of SXTs.
The first model is a non-standard TDIM because the value of the viscosity
parameter $\alpha_{\rm cold}$ used is significantly smaller than the standard
value $\sim 0.02$. The second model explores the possibility that quiescent,
truncated discs in SXTs are globally stable. This model is not a Disk
Instability Model, since the origin of an outburst has to be a variation of
the mass transfer rate in the system.

\subsection{TDIMs with a smaller $\alpha$ in quiescence}

\begin{figure}
\epsfxsize=\columnwidth
\epsfbox{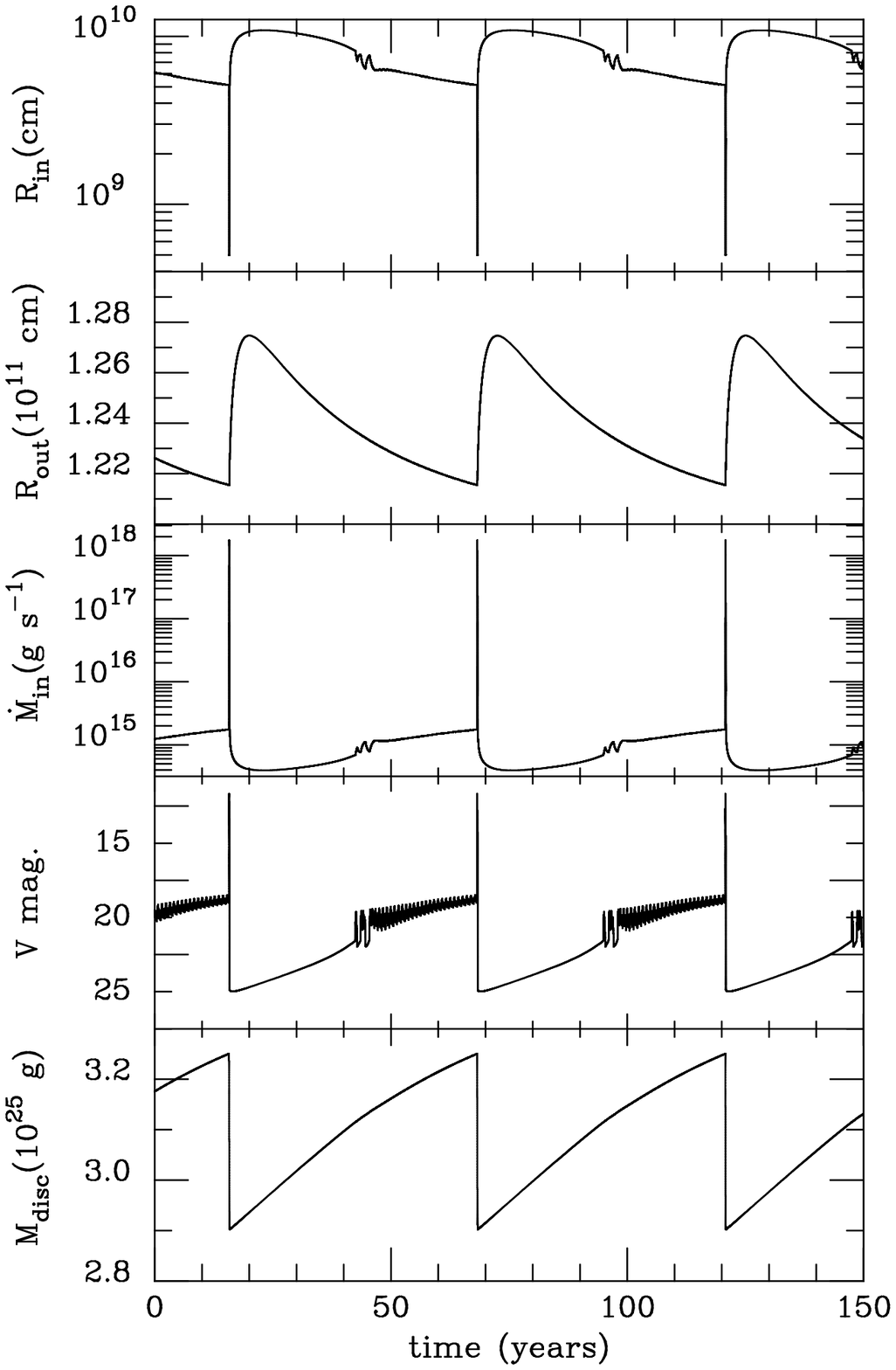}
\caption{Evolution of the disc mass $M_{\rm disc}$, the V band magnitude of
the disc, the accretion rate at the disc inner edge $\dot M_{\rm in}$, the
disc outer radius $R_{\rm out}$ and the disc inner radius $R_{\rm in}$ in a
model with parameters relevant for the BH SXT A0620-00. The recurrence time is
$t_{\rm rec} \simeq 50$ years in this model with $\alpha_{\rm cold}=5 \times
10^{-3}$.}
\label{fig:a0620lowalphacycle}
\end{figure}

\begin{figure}
\epsfxsize=\columnwidth
\epsfbox{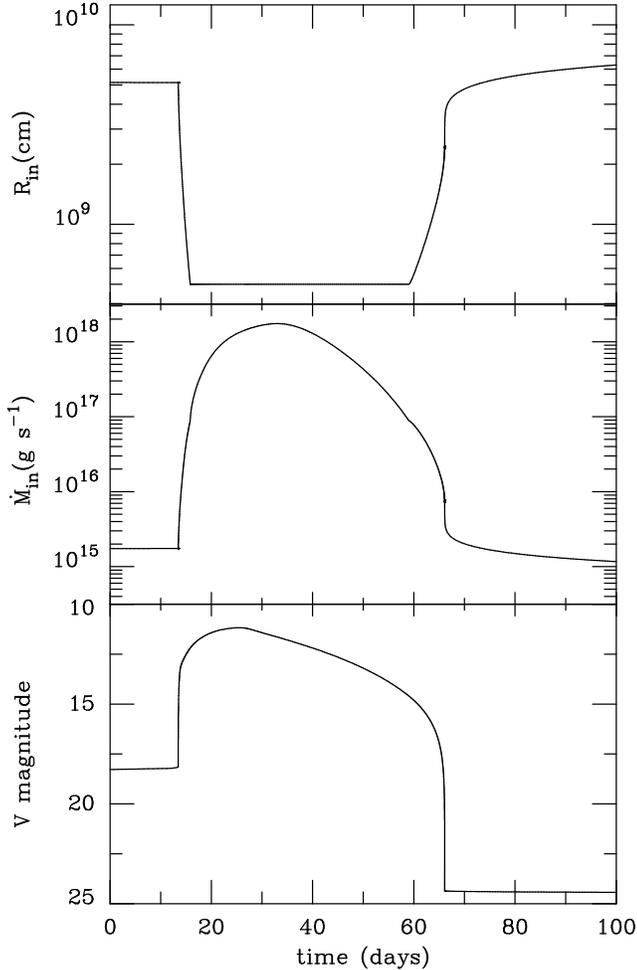}
\caption{Details on the evolution of the V band magnitude of the disc, the
accretion rate at the disc inner edge $\dot M_{\rm in}$ and the disc inner
radius $R_{\rm in}$ during an outburst of the model presented in Fig.~7}. Note
the absence of reflares.
\label{fig:a0620lowalphaoutburst}
\end{figure}

The model described in this section is similar to the model with parameters
relevant for the BH SXT A0620-00 presented in \S~3, except for $\alpha_{\rm
cold}$ which is set to $5 \times 10^{-3}$ here. We have also taken a slightly
larger ${\cal E} = 30$. The predictions for the outburst cycles are shown in
Fig.~\ref{fig:a0620lowalphacycle}, and a details on the evolution of important
quantities during an outburst are shown in
Fig.~\ref{fig:a0620lowalphaoutburst}.

As expected, the recurrence time predicted in this model is much longer
($t_{\rm rec} \sim 53$ years) and the maximum accretion rate reached in
outburst is much larger ($\sim 0.1 \dot M_{\rm Edd}$) than in the model with
$\alpha_{\rm cold}=0.02$. In our view, this is a success of the model that
such long $t_{\rm rec}$ and high $\dot M$ in outburst can be reached with
smaller values of $\alpha_{\rm cold}$.

There are other important differences caused by the reduction of $\alpha_{\rm
cold}$. The disc mass is much larger in this model, as is the total mass
accreted during an outburst. The variations of $R_{\rm out}$ are also larger.
Finally, the disc optical emission in quiescence is quite reduced compared to
what it is in the model with $\alpha_{\rm cold}=0.02$.

The most obvious difference, however, is probably the disappearance of
reflares in the decline phase of the outbursts when $\alpha_{\rm cold}=5
\times 10^{-3}$. This is in fact consistent with the explanation for the
occurrence of reflares given in \S~3.2.6. The ratio $\Sigma_{\rm
max}/\Sigma_{\rm min}$ is indeed increased from 7 to $\sim 20$ when
$\alpha_{\rm cold}$ is reduced from 0.02 to $5 \times 10^{-3}$ (for the same
$\alpha_{\rm hot}$), which makes it more difficult for the post-cooling-front
surface density to reach $\Sigma_{\rm max}$ and trigger a reflare.

The total duration of an outburst predicted in the model with $\alpha_{\rm
cold}=5 \times 10^{-3}$ is $t_{\rm dur} \simeq 50$ days (similar in optical
and X-rays). This is the same as in the case $\alpha_{\rm cold} = 0.02$ and is
still fairly short compared to the observed duration times (\S~2.1.4).

Once an outburst is triggered, it takes the disc $\sim 2$ days to reach the
compact object. The rise phase of the outburst (both in X-rays and in optical)
can be clearly separated in an early phase of very rapid increase and a late
phase of much slower increase (see Fig.~\ref{fig:a0620lowalphaoutburst}). The
total rise time for the early phase in X-rays (up to $\dot M_{\rm in} \sim 2
\times 10^{17}$~g~s$^{-1}$; see Fig.~\ref{fig:a0620lowalphaoutburst}) is
$t_{\rm rise} \simeq 3$ days, while the slower phase takes an additional $\sim
15$ days to reach the outburst peak. The total rise time for the early phase
in optical (up to $V=13$; see Fig.~\ref{fig:a0620lowalphaoutburst}) is $t_{\rm
rise} \simeq 1$ day, while the slower phase takes an additional $\sim 10$ days
to reach the outburst peak.

The rise timescales (we only consider the early phases of rapid increase here)
are $\tau_{\rm rise} \simeq 0.4$ days in X-rays (short but consistent with
observed values), and $\tau_{\rm rise} \simeq 0.1$ days in optical. The decay
timescales (early in the decline phase, before it steepens) are $\tau_{\rm
dec} \simeq 10$ days in X-rays and $\tau_{\rm dec} \simeq 13$ days in optical.
These values are too small, and not consistent with the lowest observed values.

\subsection{Models with globally stable truncated discs}

For completeness, we also constructed a model with parameters relevant for the
BH SXT A0620-00 in which the disc is truncated (same evaporation law as
before), the mass transfer rate $\dot M_T$ is low enough for the disc to be
globally stable during a long `quiescent' period, and an outburst is triggered
by a slow variation of $\dot M_T$. (Note that this is different from a burst
of mass, or an enhanced mass transfer during the outburst.) More specifically,
$\dot M_T$ was initially set to the value $10^{15}$ g s$^{-1}$ and was
linearly increased over a period of several years until it reaches a value
sufficient to trigger an outburst in the disc. The values of $\alpha$ chosen
for this model are $\alpha_{\rm hot}=0.1$ and $\alpha_{\rm cold}=0.02$, i.e.
similar to the `standard' values of the models described in \S~3.2.

We find that in most respects, the predictions of this model for the outburst
properties are similar to those of the TDIM of A0620-00 described in \S~3.2
(which is why we do not show a figure for an outburst in this model;
Fig.~\ref{fig:a0620outburst} is basically applicable): the maximum accretion
rate reached is slightly smaller, and two reflares instead of three are
predicted (as well as a duration a bit shorter).

Of course, there is no prediction for the recurrence time in this model, since
one has to refer to some activity cycle in the secondary to obtain outburst
cycles for the disc.

\section{Discussion}

We have constructed models of SXT outbursts that include evaporation of the
disc inner regions during quiescence. The only numerical limitation in our
calculations is the minimum radius $R_{\rm min}=5 \times 10^{8}$ cm allowed
for the disc inner radius $R_{\rm in}$ during evolution. Our working
assumption is that, in outburst, the disc extends down to the compact object
with a nearly constant accretion rate in its inner regions. It seems a
reasonable assumption given the results of existing calculations on such
extended discs (e.g. Cannizzo et al. 1995), and it allows us not to worry
about modeling the innermost regions of the disc during the outbursts.

One of the advantage of this simplification is that we do not have to handle
the problematic radiation-pressure dominated regions of the disc, located
close to the compact object in outburst. Our results concerning the recurrence
times should not be affected by the numerical limitation $R_{\rm in} \geq
R_{\rm min}$ since it does not affect the propagation of the cooling wave and
is irrelevant in quiescence (during which $R_{\rm in} \gg R_{\rm min}$).

There are, however, important physical effects that are not included in our
simulations. Disc irradiation by the central X-ray source is not taken into
account. Irradiation is likely to increase the duration times of the
outbursts, as well as increase the recurrence times and the maximum accretion
rates reached in outburst, as suggested by King \& Ritter (1998) and confirmed
by numerical calculations of Dubus (1999a,b). Irradiation is also likely to
increase the decay timescales, both in X-rays and optical. If the optical
emission from the system is dominated by reprocessing of X-rays by the disc
during outburst, the predicted optical light-curves could be markedly
different from those found here in models without irradiation. Nevertheless,
it is possible that irradiation is unimportant in some SXTs (e.g. Shahbaz,
Charles \& King 1998). Our results should be applicable to these systems.

Another effect that has been neglected in our simulations is the fact that the
disc could reach the 3:1 resonance during its expansion in outburst, at least
in short orbital period systems, where heating fronts can reach the outer edge
of the disc. Tidal effects could then significantly affect the outburst
properties (Osaki 1996). Similarly, enhanced mass transfer due to the
irradiation of the secondary might play an important role in SXTs, as it has
been argued for WZ Sge-type DN (Smak 1993; Hameury et al. 1997b; Augusteijn,
Kuulkers \& Shaham 1993).

One of the advantages of our calculations is that they are free of all these
complications, and isolate the effects of including the evaporation of the disc
and of reducing the value of $\alpha_{\rm cold}$. We acknowledge that none of
the models presented in this paper matches the requirements that we defined in
\S~2 for a fully successful model of SXTs. This suggests that at least one
ingredient of the model is missing (the most obvious candidate being
irradiation). In future applications of the DIM to SXTs, additional physical
effects will have to be included if ones wishes to reach a better agreement
between the model predictions and the observations.

One of the predictions of the DIM that was clearly identified in our models is
the possible occurrence of reflares during the decline from outburst. Some of
the BH SXT light-curves show `reflares' during decline from outburst but these
observed features are totally different from the reflares appearing in our
models. According to CSL, reflares (which they call `secondary maxima')
observed in X--ray transients are of three morphological types: `glitches',
`bumps' and `mini-outbursts'. Glitches are upward inflections superposed on a
smooth exponential decay. Neither the shape, nor the amplitude of reflares
obtained in our models correspond to glitches. Bumps and mini-outbursts are
more heterogeneous classes of features but it is difficult to find in the CSL
compilation a light--curve morphology similar to the one seen in
Fig.~(\ref{fig:a0620outburst}). The reflares are only present in models with
$\alpha \gg 10^{-3}$, i.e. in models with recurrence times too short to
correspond to the observations. It is, therefore, a success of the model that
the set of parameters ensuring both long recurrence times and absence of
unobserved reflares are consistent. Some of the observed reflares may be due
to disc irradiation (G. Dubus, private comm. 1999), which means that this
effect has to be included before any detailed model of SXT with reflares can
be constructed.

The fact that, in our simulations, reflares occur only when the mass of the
accreting compact object is several $M_{\odot}$ (the boundary being $\sim 3
M_{\odot}$ for standard values of $\alpha$) is interesting because it is
apparently consistent with the fact that reflares have been observed only in
SXTs containing BHs, and not in SXTs containing NSs (e.g. CSL). However, since
our reflares are different from the observed ones and absent in the favoured
model, this might be a coincidence. On the other hand, since a complete,
self-consistent model of X--ray transients is still to be constructed it could
be useful to keep in mind this property differentiating neutron star from
black holes.

By using the strongest evaporation compatible with existing upper limits on
the value of $R_{\rm in}$ in quiescent BH SXTs, we tested if evaporation alone
can be responsible for the long recurrence times of SXTs, or if very small
values of $\alpha_{\rm cold}$ in their quiescent discs are also required. Our
calculations in \S~3.2.7 show that a strong evaporation is not sufficient, and
that either a viscosity parameter $\alpha_{\rm cold}$ at the level of $\sim
10^{-3}$ is required to reach agreement with recurrence times of tens of
years, or that disc outbursts must be drastically affected by X-ray
illumination of the disc.

The very low values of $\alpha_{\rm cold}$ required to model BH SXTs would
suggest that more than one viscosity mechanism drives accretion in the disc.
The viscosity in the hot state arises probably from a well developed MHD
turbulence with a universal viscosity parameter $\alpha_{\rm hot}\gta 0.1$.
The much lower value of $\alpha$ in quiescent discs is consistent with the
disappearance of MHD turbulence in quiescent discs proposed by Gammie \& Menou
(1998) (see also Meyer \& Meyer-Hofmeister 1999). We note that if the
relevant mechanisms in quiescence were non-local in nature, the use of a local
$\alpha-$prescription would be questionable and a major revision of the DIM
might be required.

\section{Conclusion}

In this paper, we investigated the effect of disc evaporation on the
predictions of the DIM for SXTs. As for DN discs, we show that the effect of
allowing the disc outer radius $R_{\rm out}$ to vary with time is crucial to
obtain robust predictions for the outburst cycles.

We use the strongest evaporation still compatible with available upper limits
on the value of the disc inner radius in quiescent BH SXTs to test if
evaporation alone can be responsible for the long recurrence times of SXTs.

We find that such a strong evaporation does increase the recurrence times
predicted (in agreement with earlier calculations by Cannizzo 1998), but is
not sufficient to reproduce the longest known recurrence times of SXTs when
standard values of the viscosity parameter $\alpha$ are used in the disc
($\alpha_{\rm hot} \sim 0.1$, $\alpha_{\rm cold} \sim 0.02$).

We show that models with strong evaporation and a significantly smaller value
of $\alpha_{\rm cold}$ ($\sim$ a few $10^{-3}$) do reproduce the long
recurrence times of tens of years of BH SXTs, as well as the high X-ray
luminosities reached by these systems in outburst. The value of $\alpha_{\rm
cold}$ needed to reproduce the long recurrence times would be even smaller if
evaporation was weaker than what we assumed. Another possibility is of course
that X-ray illumination drastically alters the outburst cycle. We defer the
examination of this effect to a future work.

We argue that the requirement for a smaller value of $\alpha_{\rm cold}$ in at
least some SXTs is consistent with the disappearance of MHD turbulence in
quiescent discs (Gammie \& Menou 1998) and suggests that another mechanism,
perhaps non-local, could be responsible for accretion in the disc during this
low luminosity phase.

Finally, since the version of the DIM used in this paper fails to reproduce
several important properties of SXTs (such as duration and shape of the
light-curve), future work must include in the model the observed missing
ingredient: X--ray irradiation of the disc.

\section*{Acknowledgments} JPL thanks Guillaume Dubus for illuminating
discussions. This work was supported in part by the National Science
Foundation under Grant No. PHY94-07194 and by NASA grant 5-2837, and in part
by ASPS/CNRS. KM was supported by a SAO predoctoral fellowship and a French
Higher Education Ministry grant.


\begin{thebibliography}{}

\bibitem[\protect\citename{Augusteijn, Kuulkers \& Shaham }1993]{aks93}
   Augusteijn T.,  Kuulkers E., Shaham J., 1993, A\&A, 279, 13
\bibitem[\protect\citename{Cannizzo, }1993a]{c93}Cannizzo J.K., 1993a,
  ApJ, 419, 318
\bibitem[\protect\citename{Cannizzo, }1993b]{c93b}Cannizzo J.K., 1993b,
  in Wheeler J.C., ed., Accretion discs in Compact Stellar Systems. World
  Scientific, Singapore, p. 6
\bibitem[\protect\citename{Cannizzo, }1994]{c94}Cannizzo J.K., 1994,
  ApJ, 435, 389
\bibitem[\protect\citename{Cannizzo, }1998]{c98}Cannizzo J.K., 1998, ApJ, 494,
  366
\bibitem[\protect\citename{Cannizzo \& Mattei, }1992]{}Cannizzo J.K.,
  Mattei, J.A., 1992, ApJ, 401, 642
\bibitem[\protect\citename{Cannizzo, Chen \& Livio, }1995]{}Cannizzo J.K.,
  Chen W., Livio M., 1995, ApJ, 454, 880
\bibitem[\protect\citename{Cannizzo, Ghosh \& Wheeler, }1982]{cgw82}Cannizzo
  J.K., Gosh P., Wheeler J.C., 1982, ApJ, 260, L83
\bibitem[]{}Chen W., Shrader C.R., Livio M., 1997, ApJ, 491, 312
\bibitem[]{}Dubus G., 1999a, in  19th Texas Symposium on Relativistic
  Astrophysics and Cosmology, Paris, France, Dec. 14-18, in press
\bibitem[]{}Dubus G., 1999b, New AR, in press
\bibitem[]{}Dubus G., Lasota J.-P., Hameury J.-M., Charles P., 1998,
  MNRAS, 303, 139
\bibitem[\protect\citename{Esin, McClintock \& Narayan }1997]{enm97}Esin A.A.,
  McClintock J.E., Narayan R., 1997, ApJ, 489, 865
\bibitem[\protect\citename{Esin, Lasota \& Hynes }1999]{elh99}Esin A.A.,
  Lasota J.-P., Hynes, R.I., 1999, A\&A, submitted
\bibitem[\protect\citename{Esin et al. }1998]{encgz98}Esin A.A., Narayan R.,
  Cui W., Grove J.E., Shang S.-N., 1998, ApJ, 505, 854
\bibitem[\protect\citename{Gammie \& Menou, }1998]{}Gammie C.F., Menou K., 1998,
  ApJ, 492, L75
\bibitem[\protect\citename{Garcia et al. }1998]{} Garcia M. R., McClintock
  J. E., Narayan R., Callanan J., 1998, in Howell S., Kuulkers E., Woodward, C.
  eds., Wild Stars in the Old West. ASP Conf. Ser. 137, p. 506
\bibitem[\protect\citename{Hameury et al., }1997a]{hlmn97}Hameury
  J.-M., Lasota J.-P., McClintock J.E., Narayan R., 1997a, ApJ,
  489, 234
\bibitem[\protect\citename{Hameury et al. }1997b]{hlh97}Hameury
  J.-M., Lasota J.-P., Hur\'e J.-M., 1997b, MNRAS, 287, 937
\bibitem[\protect\citename{Hameury et al., }1998]{hmdlh98}Hameury
  J.-M., Menou K., Dubus G., Lasota J.-P., Hur\'e J.-M., 1998, MNRAS,
  298, 1048
\bibitem[\protect\citename{Hameury, Lasota \& Dubus }1999a]{hld99}Hameury J.-M.,
  Lasota J.-P., Dubus G., 1999a, MNRAS, 303, 39
\bibitem[\protect\citename{Hameury, Lasota \& Warner }1999b]{hlw99}Hameury
  J.-M., Lasota J.-P., Warner B., 1999b, A\&A, in press
\bibitem[\protect\citename{Honma, }1996]{}Honma F., 1996, PASJ, 48, 103
\bibitem[\protect\citename{Huang \& Wheeler, }1989]{hw89}Huang M., Wheeler
  J.C., 1989, ApJ, 343, 229
\bibitem[\protect\citename{Idan et al., }1999]{ilhs99}Idan I., Lasota J.-P.,
  Hameury J.-M., Shaviv G., 1999, Phys. Rep., 311, 213
\bibitem[\protect\citename{King \& Ritter, }
  1998]{kr98}King A.R., Ritter H., 1998, MNRAS, 293, L42
\bibitem[]{}Kuulkers E., Howell S.B., van Paradijs J., 1996, ApJ, 462,
  L87
\bibitem[]{}Kuulkers E., 1998, NewAR, 42, 1
\bibitem[\protect\citename{Lasota }1996a]{l96a}Lasota J.P., 1996a, in Evans A.,
  Wood J.H., eds., Cataclysmic Variables and Related Objects, IAU Coll. 158,
  Kluwer, Dordrecht, p. 385
\bibitem[\protect\citename{Lasota }1996a]{l96b}Lasota J.-P., 1996b, in van
  Paradijs J., van den Heuvel E.P.J., Kuulkers E., eds., Compact Stars in
  Binaries. IAU Coll. 165, Kluwer, Dordrecht, p. 43
\bibitem[\protect\citename{Lasota, }1999]{jpl99}Lasota J.-P., 1999, in
  Mineshige S., Wheeler J.C., eds., Disk Instabilities in Close Binary
  Systems -- 25 years of the Disk-Instability Model. Universal Academy Press,
  Tokyo, p. 191 (astro-ph/9901297)
\bibitem[\protect\citename{Lasota \& Hameury, }1998]{lh98}Lasota
  J.-P., Hameury J.-M., 1998, in Holt S., Kallman T., eds., Accretion
  Processes in Astrophysics - Some Like it Hot. AIP Conference Proceedings 431,
  p.351 (astro-ph/9712202)
\bibitem[\protect\citename{Lasota, Hameury \& Hur\'e }1995]{lhh95}Lasota J.-P.,
  Hameury J.-M., Hur\'e J.-M., 1995, A\&A, 302, L29
\bibitem[\protect\citename{Lasota, Narayan \& Yi }1996]{lny96}Lasota J.-P.,
  Narayan R., Yi I., 1996, A\&A, 314, 813
\bibitem[\protect\citename{Lasota, Kuulkers \& Charles }1999]{lkc99}Lasota
  J.-P., Kuulkers E., Charles P., 1999, MNRAS, 305, 473
\bibitem[\protect\citename{Liu, Meyer-Hofmeister \& Meyer, }1997]{}Liu B.F.,
  Meyer F., Meyer-Hofmeister E., 1997, A\&A, 328, 247
\bibitem[\protect\citename{Livio \& Pringle }1992]{lp92}Livio M., Pringle J.E.,
  1992, MNRAS, 229, 23P
\bibitem[\protect\citename{Mauche }1996]{m96}Mauche C.W., 1996, in Bowyer S.,
  Malina R.F., eds., Astrophysics in the Extreme Ultraviolet. IAU Coll. 152,
  Kluwer, Dordrecht, p. 317
\bibitem[\protect\citename{Menou, Hameury \& Stehle, }1999a]{mhs99}Menou K.,
  Hameury J.-M., Stehle R., 1999a, MNRAS, 305, 79
\bibitem[\protect\citename{Menou, Narayan \& Lasota, }1999b]{mnl}Menou K.,
  Narayan R., Lasota J.-P., 1999b, ApJ, 513, 811
\bibitem[\protect\citename{Meyer \& Meyer-Hofmeister }1994]{mm94}Meyer F.,
  Meyer-Hofmeister E., 1994, A\&A, 288, 175
\bibitem[\protect\citename{Meyer \& Meyer-Hofmeister }1999]{mm99}Meyer F.,
  Meyer-Hofmeister E., 1999, A\&A, 341, L23
\bibitem[\protect\citename{Mineshige \& Wheeler }1989]{mw89} Mineshige S.,
  Wheeler J.C., 1989, ApJ, 343, 241
\bibitem[\protect\citename{Narayan }1996]{n96}Narayan R., 1996, ApJ, 462, 136
\bibitem[\protect\citename{Narayan \& Yi }1995]{ny95}Narayan R., Yi I., 1995,
  ApJ, 452, 710
\bibitem[\protect\citename{Narayan, McClintock \& Yi} 1996]{nmy96} Narayan R.,
  McClintock J.E., Yi I., 1996, ApJ, 451, 821
\bibitem[\protect\citename{Narayan, Barret \& McClintock} 1997]{nbm97}
  Narayan R., Barret D., McClintock J.E., 1997, ApJ, 482, 448
\bibitem[\protect\citename{Orosz et al. }1994]{}Orosz J.A., Bailyn C.D.,
  Remillard R.A., McClintock J.E., Foltz, C.B., 1994, ApJ,  36, 848
\bibitem[\protect\citename{Orosz et al. }1997]{o97}Orosz J.A., Remillard R.A.,
  Bailyn C.D., McClintock J.E., 1997, ApJ, 478, L83
\bibitem[\protect\citename{Osaki, }1995]{o95}Osaki Y., 1995, PASP, 47, 47
\bibitem[\protect\citename{Osaki, }1996]{o96}Osaki Y., 1996, PASP,
  108, 39
\bibitem[\protect\citename{Shahbaz, Charles \& King, }1998]{sck98}Shahbaz T.,
  Charles P.A., King A.R., 1998, MNRAS, 301, 382
\bibitem[\protect\citename{Shakura \& Sunyaev, }1973]{ss73}Shakura N. I.,
  Sunyaev R.A., 1973, A\&A, 24, 337
\bibitem[\protect\citename{Shaviv, Wickramasinghe \& Wehrse, }1999]{}
  Shaviv G., Wickramasinghe D., Wehrse R., 1999, A\&A, 344, 639
\bibitem[\protect\citename{Smak, }1984]{s84}Smak J., 1984, Acta Astron., 34,
  161
\bibitem[\protect\citename{Smak, }1993]{s93}Smak J., 1993, Acta Astron., 43,
  101
\bibitem[\protect\citename{Smak, }1998]{s98}Smak J., 1998, Acta Astron., 48,
  677
\bibitem[\protect\citename{Smak, }1999]{s99}Smak J., 1999, in 
  Mineshige S., Wheeler J.C., eds., Disk Instabilities in Close Binary
  Systems -- 25 years of the Disk-Instability Model. Universal Academy Press,
  Tokyo, p. 1
\bibitem[\protect\citename{Sobczak et al. }1999]{} Sobczak G.J., McClintock 
  J.E., Remillard R.A., Bailyn C.D., Orosz J.A., 1999, ApJ, 520, 776
\bibitem[\protect\citename{Szkody \& Mattei, }1984]{szm84} Szkody P., Mattei,
  J.A., 1984, PASP, 96, 988
\bibitem[\protect\citename{Tanaka \& Lewin, }1995]{tl95}Tanaka Y., Lewin W., 
  1995, in Lewin W.H.G., van Paradijs J., van den Heuvel E.P.J., eds.,
  X-ray Binaries. Cambridge University Press, Cambridge, p. 126
\bibitem[\protect\citename{Tanaka \& Shibazaki, }1996]{}Tanaka Y.,
  \& Shibazaki N., 1996, ARA\&A, 34, 607
\bibitem[\protect\citename{van Paradijs, }1996]{}van Paradijs J., 1996, ApJ,
  464, L139
\bibitem[\protect\citename{van Paradijs \& Verbunt, }1984]{vpv84}van Paradijs
  J., Verbunt F., 1984, in Woosley S.E., ed., High Energy Transients in 
  Astrophysics, AIP Conf. Proc. 115, American Institute of Physics, New York,
  p.49
\bibitem[\protect\citename{van Paradijs \& McClintock, }1995]{vm95}
  van Paradijs J., McClintock J.E., 1995, in Lewin W.H.G., van Paradijs J.,
  van den Heuvel E.P.J., eds., X-ray Binaries. Cambridge University Press,
  Cambridge, p. 58
\bibitem[\protect\citename{Vishniac, }1997]{v97} Vishniac E.T., 1997,
  ApJ, 482, 414
\bibitem[\protect\citename{Vishniac \& Wheeler }1996]{vw96} Vishniac E.T.,
  Wheeler J.C., 1996, ApJ, 471, 921
\bibitem[\protect\citename{Warner }1995]{w95}Warner B., 1995,
  Cataclysmic Variable Stars. Cambridge University Press, Cambridge
\bibitem[\protect\citename{Warner }1998]{w98}Warner B., 1998, in Howell S.,
  Kuulkers E., Woodward C., eds., Wild Stars in the Old West. ASP Conf. Ser.
  137, p. 2
\bibitem[\protect\citename{Warner, Livio \& Tout }1996]{wlt96}Warner B., Livio
  M., Tout C.A., 1996, MNRAS, 282, 735
\bibitem[\protect\citename{White, Nagase \& Parmar, }1995]{wnp95}White N.E.,
  Nagase F., Parmar A.N., 1995, in Lewin W.H.G., van Paradijs J., van den
  Heuvel E.P.J., eds., X-ray Binaries, Cambridge University Press, Cambridge,
  p. 1
\bibitem[\protect\citename{\.Zycki, Done \& Smith, }1999]{}\.Zycki P.T.,
 Done C., Smith D.A., 1999, MNRAS, 305, 231
\end{thebibliography}
\end{document}